\documentclass[aps, prd, amsmath, floats, floatfix, twocolumn,
superscriptaddress, nofootinbib, showpacs]{revtex4-1}

\usepackage{graphicx}
\usepackage{color}
\usepackage{soul}
\usepackage{url}
\usepackage{bm}         
\usepackage{times}
\usepackage{dcolumn}
\usepackage{bm}
\usepackage{epsf}
\usepackage{amssymb}

\newcommand{\beq}{\begin{equation}}
\newcommand{\eeq}{\end{equation}}
\newcommand{\beqn}{\begin{eqnarray}}
\newcommand{\eeqn}{\end{eqnarray}}

\usepackage{color}

\newcommand{\UNH}{\affiliation {Department of Physics, University of New Hampshire, 9 Library Way, Durham NH 03824, USA}}
 
\newcommand{\RU}{\affiliation{Department of Astrophysics/IMAPP, Radboud University Nijmegen, P.O. Box 9010, 6500 GL Nijmegen, The Netherlands}}
\newcommand{\GRAPPA}{\affiliation{GRAPPA, Anton Pannekoek Institute for Astronomy and Institute of High-Energy Physics, University of Amsterdam, Science Park 904, 1098 XH Amsterdam, The Netherlands}}
\newcommand{\Nikhef}{\affiliation{Nikhef, Science Park 105, 1098 XG Amsterdam, The Netherlands}}

\begin{document}
\title{Remnant baryon mass outside of the black hole after a neutron star-black hole merger}
\author{Francois Foucart}\UNH
\author{Tanja Hinderer}\RU
\author{Samaya Nissanke}\GRAPPA \Nikhef \RU

\begin{abstract}
Gravitational-wave (GW) and electromagnetic (EM) signals from the merger of a Neutron Star (NS) and a Black Hole (BH) are a highly anticipated discovery in extreme gravity, nuclear-, and astrophysics. We develop a simple formula that distinguishes between merger outcomes and predicts the post-merger remnant mass, validated with 75 simulations. Our formula improves on existing results by describing critical unexplored regimes: comparable masses and higher BH spins. These are important to differentiate NSNS from NSBH mergers, and to infer source physics from EM signals.
\end{abstract}


\maketitle

 \textbf{\textit{Introduction}} In this new era of gravitational wave (GW) astronomy, the observation of the merger of a neutron star (NS) and black hole (BH) binary in GWs and/or electromagnetic (EM) emission remains amongst the most anticipated discoveries yet to happen~\cite{Abadie:2010}. NSBH mergers simultaneously involve strong-field gravity, supradense nuclear matter, complex microphysics, and powerful EM phenomena due to the delayed matter outflows at different timescales and frequencies. Understanding the detailed merger processes and multi-messenger signatures involved has been a longstanding challenge at the forefront of nuclear physics and astrophysics. 
 
In addition, the recent wealth of GW and EM measurements of GW170817~(\cite{TheLIGOScientific:2017qsa,GBM:2017lvd,2017ApJ...848L..13A} and references therein) indicated that the event was a binary neutron star (NSNS) merger. However, observations only allowed us to conclude definitively that: i) at least one NS was involved in the merger from ultraviolet-optical-infrared observations (e.g., \cite{2017Sci...358.1559K,2017ApJ...848L..19C,2017ApJ...848L..17C,2017Natur.551...80K,2017Sci...358.1583K,2017ApJ...848L..32M,2017ApJ...848L..18N,2017Natur.551...67P,2017Natur.551...75S,2017ApJ...848L..16S,2017ApJ...848L..27T,2017Sci...358.1565E}), and ii) the other object in the progenitor binary had a comparable mass from GW measurements~\cite{TheLIGOScientific:2017qsa,2018arXiv180511579T}. This highlights the urgent need to model both GW and EM observables of NSBH mergers, in particular in the equal-mass regime, and to identify their distinguishing features.
 
\begin{figure}
\begin{center}
\includegraphics[width=.99\columnwidth]{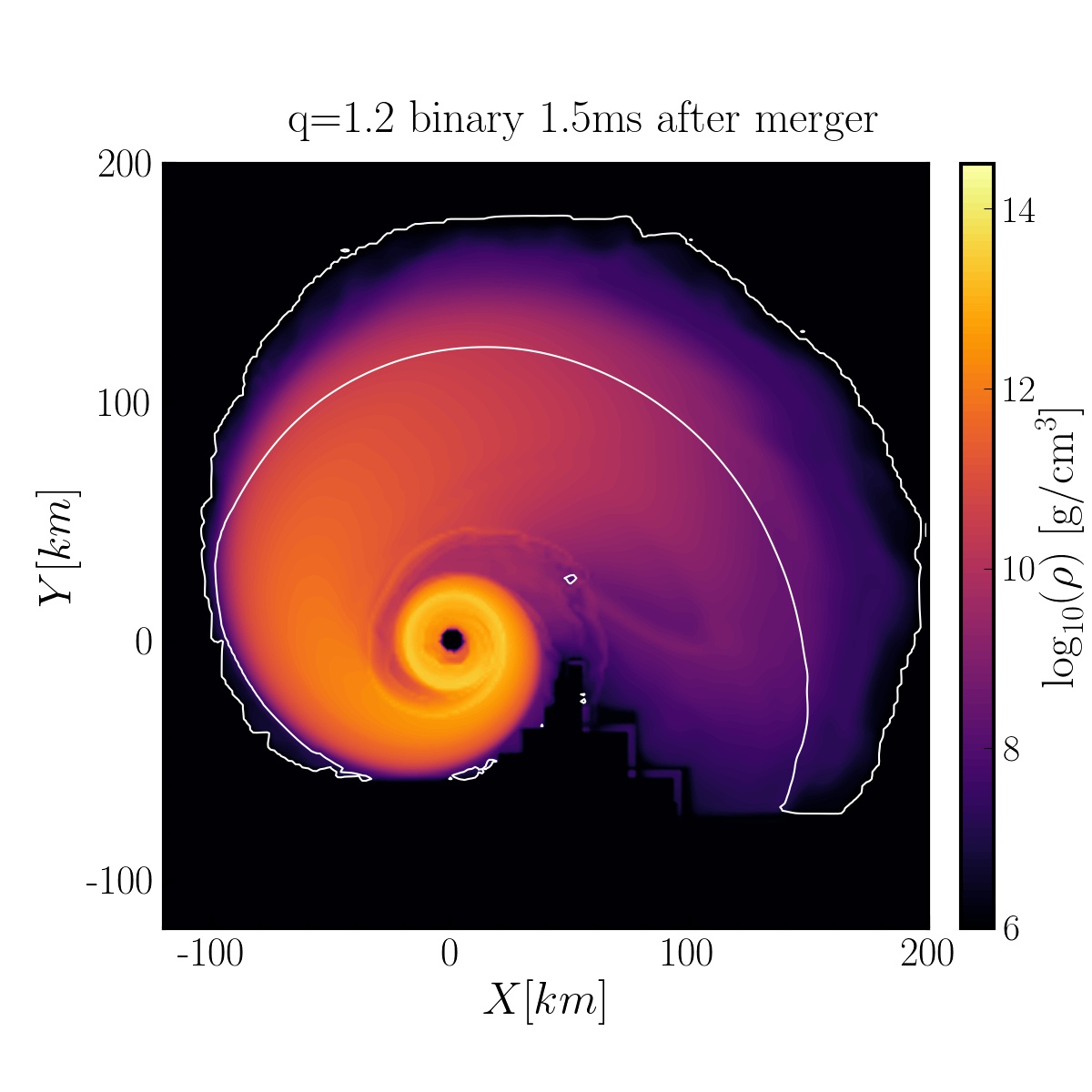}
\caption{\textit{Baryon density $1.5\,\rm ms$ after a nonspinning NSBH merger with mass ratio $Q=1.2$, in the equatorial plane.} The white contour encloses the small amount of unbound material $M_{\rm ej}\sim 5\times 10^{-4}M_\odot$. The remnant mass for such a system is significantly overestimated by the previous prediction of FF12~\cite{Foucart2012} but well-described by the new model developed here.}
\label{fig:OldFit}
\end{center}
\end{figure}

Progress on modelling the rich non-linear physics of NSBH mergers can only be obtained through numerical simulations within a general-relativistic framework. Simulations show that the NS is either torn apart by the BH's tidal forces or plunges into the BH, depending on the mass ratio, spins, and the NS Equation-of-State (EoS). If the NS is disrupted, most of the matter is accreted onto the BH within a few milliseconds. Part of the remaining material can form a disk that equilibrates after $\sim 10$ms, and bound matter in the tidal tail, illustrated in Fig.~\ref{fig:OldFit}, takes $\gtrsim 0.1$s to fall back. The merger can also eject unbound tail material. Nuclear reactions in the debris disk and ejecta, neutrino winds, and relativistic outflows are examples of processes that power EM transients such as kilonovae (e.g., \cite{1998ApJ...507L..59L,1976ApJ...210..549L,2005astro.ph.10256K,2010MNRAS.406.2650M,1998nuas.conf..103R}) and short gamma-ray bursts (e.g., \cite{1986ApJ...308L..43P,1989Natur.340..126E}). The central engine powering the latter remains a mystery.

The baryon mass outside the BH at $\sim 10\,{\rm ms}$ after merger, which we refer to as the remnant mass $M^{\rm rem}$, is therefore an important quantitative diagnostic, as shown in Foucart 2012 (hereafter FF12)~\cite{Foucart2012}. Critically, $M^{\rm rem}$ impacts the observables of the plethora of possible EM counterparts, and their detectability by current EM facilities. For instance, $M^{\rm rem}$, as introduced by FF12, is currently used when triggering EM follow-up searches by alerts sent by the LIGO and Virgo detectors (see the method outlined in \cite{Pannarale:2014rea}).  Another application is GW measurements: tidal disruption (i.e. $M^{\rm rem} \neq 0$) leads to a distinct shutoff in the signal~\cite{PhysRevD.77.084015} that depends on the long-sought after EoS of NS matter.

Previous work on modelling NSBH mergers has focused on mass ratios $Q=M_{\rm BH}/M_{\rm NS}\gtrsim 3$, with $M_{\rm BH}$ and $M_{\rm NS}$ the gravitational masses of the compact objects in isolation (see~\cite{Etienne2010,Kyutoku:2015,Kawaguchi:2015,Foucart:2014nda,Brege:2018} and references therein). This range corresponds to astrophysical formation scenarios through supernova explosions in a progenitor binary that predict a gap between NS and BH masses~\cite{Bailyn:1997xt,2010ApJ...725.1918O}. Comparable-mass binaries with a single NS could involve a primordial BH~\cite{GarciaBellido:1996qt}, a BH born in a prior NSNS merger that formed a binary through dynamical interactions in a dense cluster or galactic core (see the review in~\cite{Barack:2018yly}), or an exotic BH-like object (see \cite{Cardoso:2017njb,Barack:2018yly} for possible BH mimickers). 

In this paper we develop a simple, ready-to-use expression that predicts the range of masses, NS radii, and BH spins leading to tidal disruption, as well as $M^{\rm rem}$ for NSBH mergers. Our results cover previously unmodelled regions of the parameter space including comparable masses and high BH spins. The former are critical to distinguish NSBH from NSNS mergers. The latter are of particular interest for astrophysics and for constraining fundamental axion-like particles~\cite{Arvanitaki:2010sy}. We develop our model by deriving the dependence on the binary parameters from physical considerations about the tidal disruption process and calibrating a few added numerical coefficients to results from numerical-relativity (NR) simulations. The NR data include two novel simulations of comparable-mass binaries ($Q=1,1.2$), to be described in~\cite{NRinprep}, a case with high BH spin~\cite{Lovelace:2013vma}, and systems with a composition- and temperature-dependent EoS for the NS matter~\cite{Foucart:2014nda,Brege:2018}. 

The extreme regions in parameter space covered here are essential: we show that the model of FF12, while continuing to work well within its expected range of validity, leads to a substantially inaccurate estimate for tidal disruption and the amount of remnant mass for binary parameters outside of that range. Specifically, the remnant mass is shown to be significantly lower for nearly equal-mass NSBH mergers and higher for large BH spins than previously predicted by FF12. We also discuss two important applications of our results as examples of their utility: (i) verifying the reliability of disk mass predictions by comparing different NR codes, which is a pressing open problem that has not yet been addressed for NSBH mergers, and (ii) deriving the range of binary parameters that lead to tidal disruption, which determines if the binary is likely to have a shutoff GW signature and an EM counterpart.

\smallskip

\textbf{\textit{Numerical simulations}} We consider results from 75 NR simulations performed with three different evolution codes compiled from ~\cite{FoucartEtAl:2011,Foucart:2010eq,Kyutoku:2011vz,Etienne:2008re,Kyutoku:2011vz,Lovelace:2013vma,Foucart:2013a,Brege:2018,Foucart:2014nda,Kyutoku:2015} (see also supplementary material Table~\ref{tab:NRdataOld} and~\ref{tab:NRdataNew}).
Each simulation is parameterized by three dimensionless quantities: the mass ratio $Q\geq 1$, the dimensionless BH spin $\chi_{\rm BH}=c |{\bm S}|/(G M_{\rm BH}^2)$, where ${\bm S}$ is the spin angular momentum, and the NS's compaction $C_{\rm NS} = GM_{\rm NS}/(R_{\rm NS}c^2)$, where $R_{\rm NS}$ is the NS's areal radius that depends on the EoS. Effects of precession, NS spin, orbital eccentricity, and magnetic fields are not considered here. The simulations range over $Q\in [1,7]$, $\chi_{\rm BH}\in[-0.5, 0.97]$, and $C_{\rm NS}\in [0.13,0.182]$ and include  44 systems not used in FF12, with 11 cases having a $Q$ or $M^{\rm rem}$ outside the range of validity of FF12, and 12 systems with tabulated composition- and temperature-dependent EoS. We focus on the normalized remnant mass \beq 
\hat{M}^{\rm rem}=M^{\rm rem}/M^b_{\rm NS},
\eeq 
where $M^b_{\rm NS}$ is the baryonic mass of the initial NS. Since most of the simulation results do not have error bars, we estimate the errors $\sigma_{\rm NR}$ in $\hat{M}^{\rm rem}$ based on a few simulations where well-determined errors were computed (see FF12 for details). The resulting error estimate combines a $10\%$ relative error and a $1\%$ absolute error in the mass measurements:  
\beq
\sigma_{\rm NR} = \left[ \left(\frac{\hat{M}_{\rm rem, NR}}{10} \right)^2 + \left(\frac{1}{100} \right)^2 \right]^{1/2}.
\label{eq:sigNR}
\eeq

\smallskip

\textbf{\textit{Model for the remnant baryon mass}} 
We begin constructing our model for $\hat{M}^{\rm rem}$ with physical insights about tidal disruption. For $Q\to \infty$, the NS is tidally disrupted if it overflows its Roche lobe at a binary separation greater than the radius of the innermost stable circular orbit (ISCO) of the BH, where its motion transitions from an inspiral to a rapid plunge. In Newtonian gravity, the disruption separation is $d_{\rm dis}=(3Q)^{1/3}R_{\rm NS}$. The normalized ISCO radius $\hat{R}_{\rm ISCO} = R_{\rm ISCO}/M_{\rm BH}$ is, for $Q\rightarrow \infty$,
\beq 
\hat{R}_{\rm ISCO} = 3+Z_2-{\rm sgn}(\chi_{\rm BH})\sqrt{(3-Z_1)(3+Z_1+2Z_2)},\qquad
\eeq
with $ Z_1=1+(1-\chi_{\rm BH}^2)^{1/3}[(1+\chi_{\rm BH})^{1/3}+(1-\chi_{\rm BH})^{1/3}]$ and $Z_2=\sqrt{3\chi_{\rm BH}^2+Z_1^2}$~\cite{Bardeen1972}. The ratio of $R_{\rm ISCO}$ to the NS radius can be expressed as $R_{\rm ISCO}/R_{\rm NS}=\hat{R}_{\rm ISCO}\, Q \,C_{\rm NS}$. However, we note that $Q$ is not an ideal parameter to use when extrapolating results obtained in the high-$Q$ regime to the $Q\sim 1$ regime. This is because the $Q\to\infty$ limit obscures a symmetry of many observables in the general relativistic two-body problem: invariance under exchanging the bodies' labels. In post-Newtonian expansions this property becomes explicit: many quantities depend on the symmetric mass ratio $\eta=Q/(1+Q)^2$, which is equivalent to $Q^{-1}$ for $Q\to \infty$. When restoring this symmetry by replacing $Q$ by $\eta^{-1}$, results derived in the large-$Q$ limit can often be surprisingly accurate when compared to NR data for $Q\sim 1$ (see e.g.~\cite{Buonanno:1998gg,LeTiec-Mroue:2011}). Hence, we replace $Q^{-1}$ by $\eta$ in our considerations of tidal disruption and propose the model
\beq
\hat{M}^{\rm rem}_{\rm model} = \left[{\rm Max}\left(\alpha \frac{1-2C_{\rm NS}}{\eta^{1/3}} - \beta \hat{R}_{\rm ISCO}\frac{C_{\rm NS}}{\eta}+\gamma,0\right)\right]^\delta
\label{eq:newfit}
\eeq
with free parameters $(\alpha,\beta,\gamma,\delta)$. The first term is proportional to $d_{\rm dis}/R_{\rm NS}$, multiplied by $(1-2C_{\rm NS})$ to account for the fact that a BH (having an effective $C_{\rm BH}=1/2$ when nonspinning) cannot be tidally disrupted. The second term scales as $R_{\rm ISCO}/R_{\rm NS}$ as $Q\rightarrow \infty$. The parameters $(\gamma,\delta)$ represent nonlinear effects not accounted for by the simple physical considerations. A zero $\hat{M}^{\rm rem}_{\rm model}$ corresponds to no tidal disruption. 

\begin{figure}
\includegraphics[width=.99\columnwidth]{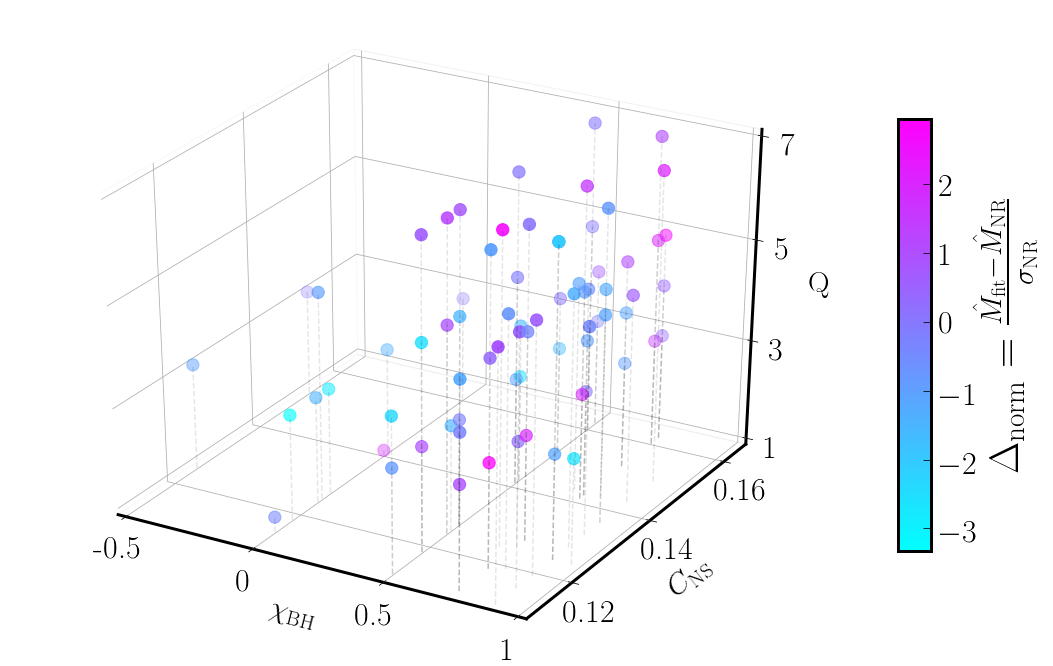}
\caption{\textit{Differences between NR results and the model}~\eqref{eq:newfit} weighted by the estimated NR error as a function of the mass ratio, NS compaction, and BH spin. Magenta (cyan) color corresponds to an over-(under-) estimate of $\hat{M}_{\rm rem}$.}
\label{fig:ErrorVsParams}
\end{figure}

To determine the free parameters in~\eqref{eq:newfit}, we first define a normalized error $\Delta_{\rm norm}$ as the difference between $\hat{M}^{\rm rem}$ computed from a model and the NR result, relative to the NR error:
\beq
\Delta_{\rm norm} = \frac{\hat{M}^{\rm rem}_{\rm  model}-\hat{M}^{\rm rem}_{\rm NR}}{\sigma_{\rm NR}}.
\label{eq:Dn}
\eeq
Minimizing the root-mean-square of $\Delta_{\rm norm}$ leads to
\beq
\alpha=0.406, \; \; \beta=0.139,\; \; \gamma=0.255, \; \; \delta=1.761.\; \; 
\label{eq:fitparameters}
\eeq
The root-mean-square error in the model is $\Delta_{\rm norm} \sim 1.4 $, and Figure~\ref{fig:ErrorVsParams} illustrates that it performs well across the 3D binary parameter space covered by simulations. 

\begin{figure}
\begin{center}
\includegraphics[width=.99\columnwidth]{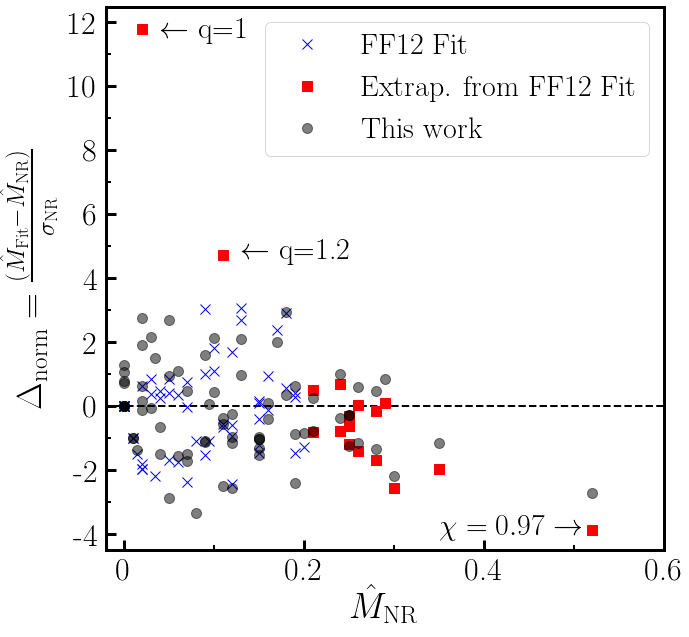}
\caption{
\textit{Normalized errors in the remnant mass predictions versus NR results.} Blue crosses indicate results for FF12 used within its range of validity, red squares for FF12 extrapolated outside that range, and grey circles for all cases, using our new model~(\ref{eq:newfit}).}
\label{fig:FitPerf}
\end{center}
\end{figure}

Our model will critically help differentiate NSNS from low-mass NSBH mergers, requiring our prediction remains robust for $Q\sim 1$ binaries. Figure~\ref{fig:FitPerf} shows that this is not the case for the model of FF12, which was derived for a narrower range of parameters ($Q\ge 3$, $\hat{M}_{\rm rem}\le 0.2$). FF12 continues to work well within that range, but substantially overestimates $M^{\rm rem}$ for $Q\sim 1$. Experimenting with our model we find that the vast improvement over FF12 for $Q\sim 1$ comes from 
the substitution of $\eta$ for $Q^{-1}$ in~(\ref{eq:newfit}), which was not done in FF12. As a result, for $Q\sim 1$, the ISCO-term trends to larger values here than in FF12 and we predict that more material falls into the BH. A larger ISCO for small $Q$ than computed for $Q\rightarrow \infty$ is expected on physical grounds: (i) the NS's plunge, although not well-defined in this limit, would begin at an effective ISCO of the two-body spacetime, and (ii) the NS matter accreted at merger causes the BH to grow, which moves the ISCO for the remaining material outwards~\cite{KK:private:2018}. The latter leads to a larger fractional change in the ISCO location for smaller $Q$. We verify that our result is not overly dependent on the 2 simulations with $Q<2$ by re-fitting~\eqref{eq:newfit} but ignoring simulations with $Q<2$. We find that this modified fit is as consistent with the low-$Q$ NR simulations as~(\ref{eq:newfit},\ref{eq:fitparameters}), obtained using all NR results. Conversely, re-fitting FF12 using all simulations still provides poor predictions for $Q\sim 1$.

Cases with large remnant masses or, equivalently, large BH spins are another interesting regime for multimessenger observations and fundamental physics that are not well-modelled by FF12. For capturing the merger outcomes in this limit, it is necessary to introduce the nonlinearity parameter $\delta$, which was not included in FF12. 

\begin{figure}
\begin{center}
\includegraphics[width=.99\columnwidth]{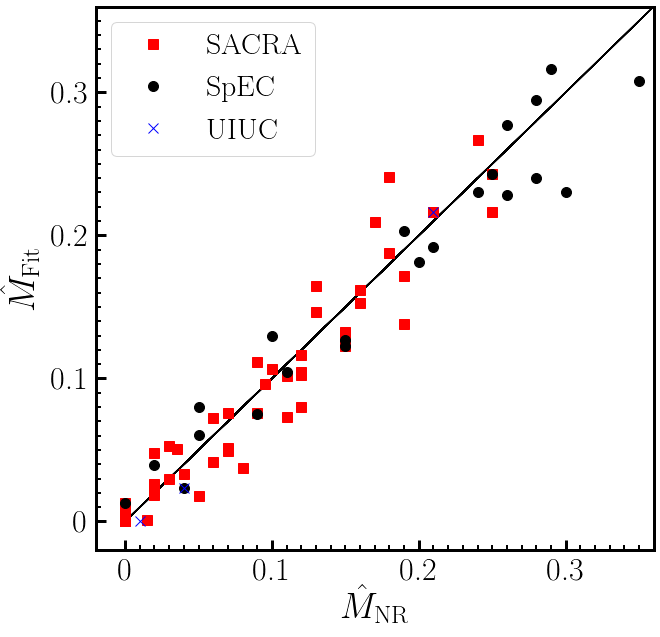}
\caption{\textit{Remnant baryon masses for all simulations used in this work, split by evolution code:} SACRA (red squares), SpEC (black circles), and the UIUC code (blue crosses). Different codes measure remnant baryon masses to within the model's accuracy.}
\label{fig:SACRAvsSPEC}
\end{center}
\end{figure}

We also verify that different NR codes predict broadly consistent $M^{\rm rem}$. Figure~\ref{fig:SACRAvsSPEC} summarizes our findings: over the range of binary parameters for which data from multiple collaborations is available, there is no systematic bias associated with the NR code used. While this comparison is not as direct as one based on identical initial data, it advantageously uses a large number of numerical results rather than comparing isolated examples, and increases our confidence in the reliability of NR simulations to within the accuracy of our model.

\begin{figure*}
\begin{center}
\includegraphics[width=.99\textwidth]{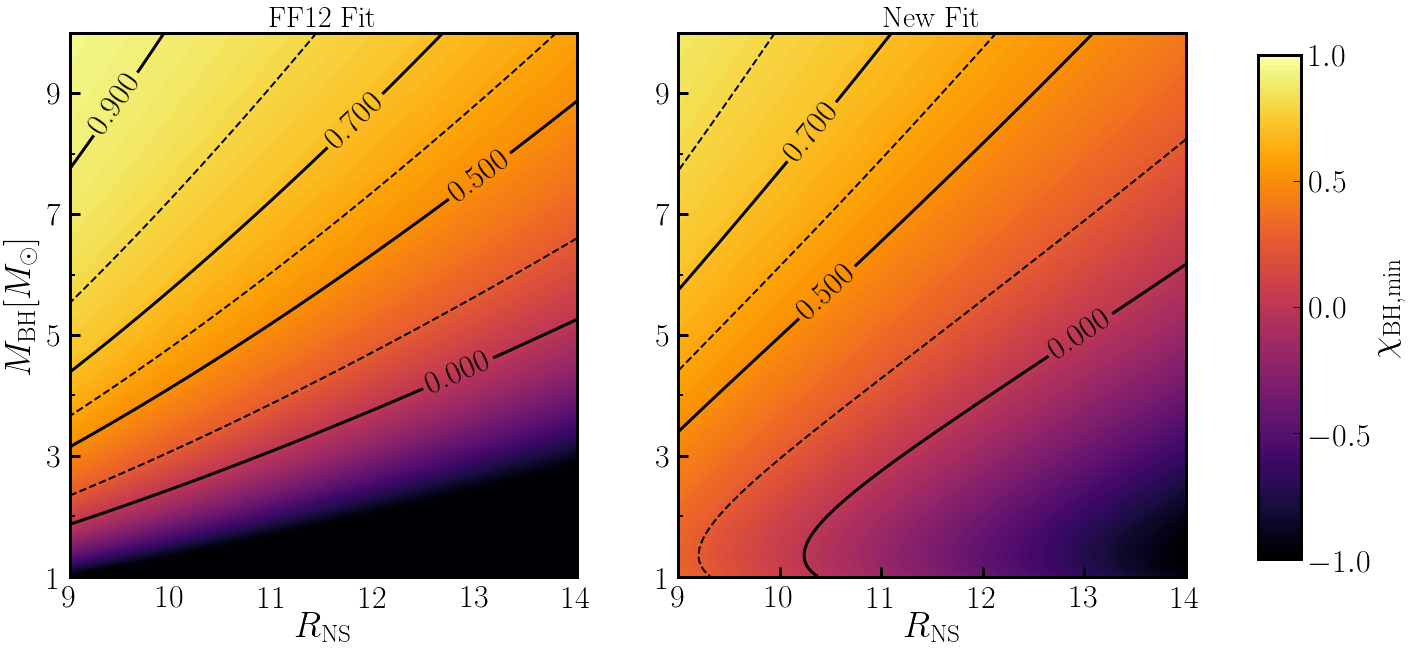}
\caption{\textit{Minimum aligned component of the BH spin required for a $1.35M_\odot$ NS to disrupt, as a function of NS radius and BH mass.} We show results for the fit from FF12 (left), and the updated formula presented here (right). Solid black curves are contours of constant $\chi_{\rm BH,min}=0,0.5,0.7,0.9$, while dashed black curves are for $\chi_{\rm BH,min}=0.25,0.6,0.8$. The updated results are less favorable to disruption for near-equal mass systems, and more favorable for large mass ratios.}
\label{fig:ChiMinDisrupt}
\end{center}
\end{figure*}

We further study an alternative model that depends on the properties of NS matter through the dimensionless quadrupolar tidal deformability of the NS, $\Lambda = (2/3) k_2 C_{\rm NS}^{-5}$ (with $k_2$ the tidal Love number), which is the best-measured EoS parameter of a slowly-spinning NS from GW data~\cite{TheLIGOScientific:2017qsa,2018arXiv180511579T,2018arXiv180511581T}. Defining $\rho = (15 \Lambda)^{-1/5}$, so that $\rho \approx C_{\rm NS}$, we use
\beq
\hat{M}^{\rm rem}_{ \Lambda}  = \left[{\rm Max}\left(\alpha \frac{1-2\rho}{\eta^{-1/3}} - \beta \hat{R}_{\rm ISCO}\frac{\rho}{\eta}+\gamma,0\right)\right]^\delta,
\label{eq:Lambdafit}
\eeq
with the best-fit parameters $\alpha=0.308$, $\beta=0.124$, $\gamma=0.283$, and $\delta=1.536$. This prediction performs as well as the compaction-based model~\eqref{eq:newfit} for $M_{\rm rem}\leq 0.2M_\odot$, but has larger errors for simulations using single-polytrope equations of state and producing more massive remnants. We find that using approximately universal relations for $C(\Lambda)$ ~\cite{Yagi:2016bkt,Maselli:2013mva} in~\eqref{eq:newfit} performs similarly to~\eqref{eq:Lambdafit}. Larger errors for single polytropes are then expected, as the properties of NSs for single-polytrope EoSs are not well-described by universal relations. While~\eqref{eq:newfit} is a priori preferable, the $\Lambda$-based model captures well the limit between disrupting and non-disrupting systems, is advantageous for rapid GW analysis, and may only be less accurate for unphysical EoSs.

\textbf{\textit{Discussion}} A key application of the new model for the remnant mass outside the BH, \eqref{eq:newfit} and \eqref{eq:fitparameters}, is to derive limits on the range of binary parameters leading to the disruption of a NS, as shown in Fig.~\ref{fig:ChiMinDisrupt}. 
For the previously unexplored $Q\sim 1$ systems, the disk masses are significantly lower than predicted by FF12, and some compact neutron stars can entirely avoid disruption. Accordingly, {\it FF12 should not be used to assess whether an observed NSNS merger could instead be a NSBH merger}. We discuss the implications of our results for the pressing question of whether GW170817 could have been a NSBH system in a companion paper. For large $Q$, our updated model is mildly more favorable to disruption than FF12.

This paper provides a simple formula for computationally inexpensive yet reliable estimates of $M_{\rm rem}$ for NSBH mergers across a wide range of parameters, including interesting regimes not previously considered. The average relative error in the remnant mass prediction is $\sim 15\%$ for binaries with $ Q\in [1, 7]$, $ \chi_{\rm BH} \in [-0.5, 0.9]$, and $M_{\rm rem}\lesssim 0.3 M^b_{\rm NS}$. Our work can guide the future choice of progenitor parameters for NR simulations, and detailed models of EM counterparts and GW emission, a requisite for identifying and characterizing NSBH mergers.

\smallskip

\textbf{\textit{Acknowledgments}}
We thank Kenta Hotokezaka, Koutarou Kyutoku, Kyohei Kawaguchi, Shaon Ghosh, David Nichols, Patricia Schmidt, Andrew Williamson, and the members of the SxS collaboration for useful discussions and comments. FF gratefully acknowledges support from NASA through grant 80NSSC18K0565. TH is grateful for financial support through the Radboud University Excellence fellowship scheme. SMN is grateful for support from NWO VIDI and TOP Grants of the Innovational Research Incentives Scheme (Vernieuwingsimpuls) financed by the Netherlands Organization for Scientific Research (NWO).

\bibliographystyle{iopart-num}
\bibliography{References.bib}

\appendix

\begin{table}
\begin{center}
\caption{Numerical-relativity simulations used in our fits that were already used in FF12. 
We list the mass ratio $Q=M_{\rm BH}/M_{\rm NS}$, the neutron star
compaction $C_{\rm NS}=GM_{\rm NS}/(R_{\rm NS}c^2)$, the dimensionless spin-parameter of the black hole $\chi_{\rm BH}$, the parameter $\rho=(15\Lambda)^{1/5}$ used in Eq.~\ref{eq:Lambdafit} (with $\Lambda$ the dimensionless tidal deformability of the neutron star), and the baryon mass remaining outside
of the black hole $10\,{\rm ms}$ after merger, $M^{\rm rem}$, normalized to initial baryon mass of the neutron star $M^b_{\rm NS}$. We also provide the type of equation of state, the code used to perform the simulation, and a reference to the relevant publication. The $\Gamma 2$ equation of state denotes an ideal gas with polytropic index $\Gamma=2$, while PP refers to piecewise polytropic equations of state.}
\begin{tabular}{c|c|c|c|c|c|c|c}
$Q$ & $\chi_{\rm BH}$ & $C_{\rm NS}$  & $\rho$ & $\frac{M^{\rm rem }}{M^b_{\rm NS}}$  & EoS Type & Code & Reference \\
\hline
7.0  &  0.9  &  0.144  &  6.56 & 0.24  & $\Gamma 2$ & SpEC & \cite{FoucartEtAl:2011}\\
7.0  &  0.7  &  0.144  &  6.56 &  0.05  & $\Gamma 2$ & SpEC & \cite{FoucartEtAl:2011}\\
5.0  &  0.5  &  0.144  &  6.56 &  0.05  & $\Gamma 2$ & SpEC & \cite{FoucartEtAl:2011}\\
3.0  &  0.9  &  0.144  &  6.56 &  0.35  & $\Gamma 2$ & SpEC & \cite{Foucart:2010eq} \\
3.0  &  0.5  &  0.145  &  6.51 &  0.15  & $\Gamma 2$/PP & SpEC/SACRA & \cite{Foucart:2010eq,Kyutoku:2011vz} \\
3.0  &  0.0  &  0.144  &  6.56 &  0.04  & $\Gamma 2$ & UIUC/SpEC & \cite{Foucart:2010eq,Etienne:2008re} \\
3.0  &  0.75  &  0.145  &  6.51 &  0.21  & $\Gamma 2$/PP & UIUC/SACRA & \cite{Etienne:2008re,Kyutoku:2011vz}\\
5.0  &  0.75  &  0.131  &  8.10 &  0.25  & PP & SACRA & \cite{Kyutoku:2011vz}\\
5.0  &  0.75  &  0.162  &  6.21 &  0.11  & PP& SACRA & \cite{Kyutoku:2011vz}\\
5.0  &  0.75  &  0.172  &  5.75 &  0.06  & PP& SACRA & \cite{Kyutoku:2011vz}\\
5.0  &  0.75  &  0.182  &  5.33 &  0.02  & PP& SACRA & \cite{Kyutoku:2011vz}\\
4.0  &  0.75  &  0.131  &  8.10 &  0.25  & PP& SACRA & \cite{Kyutoku:2011vz}\\
4.0  &  0.75  &  0.162  &  6.21 &  0.15  & PP& SACRA & \cite{Kyutoku:2011vz}\\
4.0  &  0.75  &  0.172  &  5.75 &  0.12  & PP& SACRA & \cite{Kyutoku:2011vz}\\
4.0  &  0.75  &  0.182  &  5.33 &  0.07  & PP& SACRA & \cite{Kyutoku:2011vz}\\
4.0  &  0.5  &  0.131  &  8.10 &  0.19  & PP & SACRA & \cite{Kyutoku:2011vz}\\
4.0  &  0.5  &  0.162  &  6.21 &  0.06  & PP & SACRA & \cite{Kyutoku:2011vz}\\
4.0  &  0.5  &  0.172  &  5.75 &  0.02  & PP& SACRA & \cite{Kyutoku:2011vz}\\
3.0  &  0.75  &  0.131  &  8.10 &  0.24  & PP & SACRA & \cite{Kyutoku:2011vz}\\
3.0  &  0.75  &  0.162  &  6.21 &  0.16  & PP & SACRA & \cite{Kyutoku:2011vz}\\
3.0  &  0.75  &  0.172  &  5.75 &  0.15  & PP & SACRA & \cite{Kyutoku:2011vz}\\
3.0  &  0.75  &  0.182  &  5.33 &  0.1  & PP & SACRA & \cite{Kyutoku:2011vz}\\
3.0  &  0.5  &  0.131  &  8.10 &  0.19  & PP & SACRA & \cite{Kyutoku:2011vz}\\
3.0  &  0.5  &  0.162  &  6.21 &  0.11  & PP & SACRA & \cite{Kyutoku:2011vz}\\
3.0  &  0.5  &  0.172  &  5.75 &  0.07  & PP & SACRA & \cite{Kyutoku:2011vz}\\
3.0  &  0.5  &  0.182  &  5.33 &  0.03  & PP & SACRA & \cite{Kyutoku:2011vz}\\
7.0  &  0.5  &  0.144  &  6.56 &  0.0  & PP & SpEC  & \cite{FoucartEtAl:2011}\\
3.0  &  -0.5  &  0.145  &  6.51 &  0.01  & PP & UIUC & \cite{Etienne:2008re}\\
5.0  &  0.0  &  0.145  &  6.51 &  0.01  & PP & UIUC & \cite{Etienne:2008re}\\
4.0  &  0.5  &  0.182  &  5.33 &  0.0  & PP & SACRA & \cite{Kyutoku:2011vz}\\
3.0  &  -0.5  &  0.172  &  5.75 &  0.0  & PP & SACRA & \cite{Kyutoku:2011vz}\\
\end{tabular}
\label{tab:NRdataOld}
\end{center}
\end{table}

\begin{table}
\begin{center}
\caption{Same as Table~\ref{tab:NRdataOld}, but for simulations that were not used in FF12. The first 11 simulations are outside of the range of parameters covered by the fit derived in FF12. {\it Tab} refers to tabulated, composition and temperature dependent equations of state.}
\begin{tabular}{c|c|c|c|c|c|c|c}
$Q$ & $\chi_{\rm BH}$ & $C_{\rm NS}$  & $\rho$ & $\frac{M^{\rm rem }}{M^b_{\rm NS}}$  & EoS Type & Code & Reference \\
\hline
1.0  &  0.0  &  0.16  & 6.19 &  0.02  & Tab & SpEC & In Prep.\\
1.2  &  0.0  &  0.134  & 7.58 &  0.11  & Tab & SpEC & In Prep.\\
3.0  &  0.97  &  0.144  &  6.56 &  0.52  & $\Gamma 2$ & SpEC & \cite{Lovelace:2013vma}\\
7.0  &  0.9  &  0.144  &  6.56 &  0.3  & $\Gamma 2$ & SpEC & \cite{Foucart:2013a} \\
5.83  &  0.9  &  0.135  &  7.52 &  0.28  & Tab  & SpEC & \cite{Brege:2018} \\
5.0  &  0.9  &  0.156  &  6.38 &  0.26  & Tab & SpEC & \cite{Brege:2018} \\
5.83  &  0.9  &  0.13  &  7.77 &  0.29  & Tab & SpEC & \cite{Brege:2018} \\
5.0  &  0.9  &  0.152  &  6.51 &  0.25  & Tab & SpEC & \cite{Brege:2018} \\
5.83  &  0.9  &  0.148  & 6.66 &  0.28  & Tab & SpEC & \cite{Brege:2018}  \\
5.83  &  0.9  &  0.139  &  7.73 &  0.26  & Tab & SpEC & \cite{Foucart:2014nda}\\
5.83  &  0.8  &  0.139  &  7.73 &  0.21  & Tab & SpEC & \cite{Foucart:2014nda} \\
7.0  &  0.9  &  0.156  &  5.90 &  0.2  & $\Gamma 2$& SpEC & \cite{Foucart:2013a}\\
7.0  &  0.9  &  0.17  &  5.24 &  0.1  & $\Gamma 2$& SpEC & \cite{Foucart:2013a} \\
5.0  &  0.7  &  0.163  &  6.30 &  0.09  & Tab & SpEC & \cite{Foucart:2014nda}\\
5.0  &  0.8  &  0.163  &  6.30 &  0.15  & Tab & SpEC & \cite{Foucart:2014nda}\\
5.0  &  0.9  &  0.163  &  6.30 &  0.19  & Tab & SpEC & \cite{Foucart:2014nda}\\
3.0  &  0.75  &  0.18  &  5.45 &  0.09  & PP & SACRA & \cite{Kyutoku:2015}\\
3.0  &  0.75  &  0.161  &  6.44 &  0.13  & PP& SACRA & \cite{Kyutoku:2015}\\
3.0  &  0.75  &  0.147  &  7.00 &  0.17  & PP& SACRA & \cite{Kyutoku:2015}\\
3.0  &  0.75  &  0.138  &  7.66 &  0.18  & PP& SACRA & \cite{Kyutoku:2015}\\
3.0  &  0.5  &  0.18  &  5.45 &  0.04  & PP& SACRA & \cite{Kyutoku:2015}\\
3.0  &  0.5  &  0.161  &  6.44 &  0.09  & PP& SACRA & \cite{Kyutoku:2015}\\
3.0  &  0.5  &  0.147  &  7.00 &  0.12  & PP& SACRA & \cite{Kyutoku:2015}\\
3.0  &  0.5  &  0.138  &  7.66 &  0.13  & PP& SACRA & \cite{Kyutoku:2015}\\
3.0  &  0.0  &  0.18  &  5.45 &  0.0  & PP& SACRA & \cite{Kyutoku:2015}\\
3.0  &  0.0  &  0.161  &  6.44 &  0.015  & PP& SACRA & \cite{Kyutoku:2015}\\
3.0  &  0.0  &  0.147  &  7.00 &  0.05  & PP& SACRA & \cite{Kyutoku:2015}\\
3.0  &  0.0  &  0.138  &  7.66 &  0.08  & PP& SACRA & \cite{Kyutoku:2015}\\
5.0  &  0.75  &  0.18  &  5.45 &  0.03  & PP& SACRA & \cite{Kyutoku:2015}\\
5.0  &  0.75  &  0.161  &  6.44 &  0.12  & PP& SACRA & \cite{Kyutoku:2015}\\
5.0  &  0.75  &  0.147  &  7.00 &  0.16  & PP& SACRA & \cite{Kyutoku:2015}\\
5.0  &  0.75  &  0.138  &  7.66 &  0.18  & PP& SACRA & \cite{Kyutoku:2015}\\
5.0  &  0.5  &  0.18  &  5.45 &  0.0  & PP& SACRA & \cite{Kyutoku:2015}\\
5.0  &  0.5  &  0.161  &  6.44 &  0.02  & PP& SACRA & \cite{Kyutoku:2015}\\
5.0  &  0.5  &  0.147  &  7.00 &  0.07  & PP& SACRA & \cite{Kyutoku:2015}\\
5.0  &  0.5  &  0.138  &  7.66 &  0.12  & PP& SACRA & \cite{Kyutoku:2015}\\
7.0  &  0.75  &  0.18  &  5.45 &  0.0  & PP& SACRA & \cite{Kyutoku:2015}\\
7.0  &  0.75  &  0.161  &  6.44 &  0.035  & PP& SACRA & \cite{Kyutoku:2015}\\
7.0  &  0.75  &  0.147  &  7.00 &  0.095  & PP& SACRA & \cite{Kyutoku:2015}\\
7.0  &  0.75  &  0.138  &  7.66 &  0.15  & PP& SACRA & \cite{Kyutoku:2015}\\
7.0  &  0.5  &  0.18  &  5.45 &  0.0  & PP& SACRA & \cite{Kyutoku:2015}\\
7.0  &  0.5  &  0.161  &  6.44 &  0.0  & PP& SACRA & \cite{Kyutoku:2015}\\
7.0  &  0.5  &  0.147  &  7.00 &  0.0 & PP & SACRA & \cite{Kyutoku:2015}\\
7.0  &  0.5  &  0.138  &  7.66 &  0.02 & PP & SACRA & \cite{Kyutoku:2015} 
\end{tabular}
\label{tab:NRdataNew}
\end{center}
\end{table}

\end{document}